\documentclass[fp,twocolumn]{jpsj3}
\usepackage{txfonts}
\usepackage{bm}
\usepackage{color}
\usepackage[dvipsnames]{xcolor}

\title{Drag of Two Cylindrical Intruders in a Two-Dimensional Granular Environment}

\author{Takumi Kubota$^1$, Haruto Ishikawa$^1$, and Satoshi Takada$^{1,2}$\thanks{Corresponding author. E-mail: takada@go.tuat.ac.jp}}
\inst{$^1$Department of Mechanical Systems Engineering, 
Tokyo University of Agriculture and Technology,
Koganei, Tokyo 184--8588, Japan \\
$^2$Institute of Engineering, 
Tokyo University of Agriculture and Technology, 
Koganei, Tokyo 184--8588, Japan}

\abst{
	The drag of two cylindrical intruders in a two-dimensional granular environment is numerically studied by the discrete element method. 
	We find the yield force, below which the intruders cannot move because of interactions with the surrounding particles. 
	Above the yield force, on the other hand, the intruders can move at a constant speed.
	We investigate the relationship between the drag force and the steady speed of the intruders, where the speed becomes higher as the distance between the intruders decreases. 
	We confirm that the origin of the yield is the Coulombic friction between the particles and the bottom plate by changing the value of the friction coefficient. 
	We also find that the yield force is almost proportional to the friction coefficient, which means that the number of particles determining the yield force is almost constant. 
	On the other hand, the two-dimensional elasticity is applicable to determine the stress fields around the intruders. 
	We confirm that fields asymmetric with respect to the drag direction are reproduced by using the information of the stresses on the surfaces of the intruders by introducing bipolar coordinates.
}

\begin{document}
\maketitle

\section{Introduction}\label{sec:intro}

An understanding of the resistance to intruders into a flow is important in many applications \cite{Lamb,Batchelor,Takada20jet}.
The drag in a fluid is one of the principal problems that has been studied for a long time \cite{Batchelor}, where the drag is proportional to the speed or the square of the speed if the flow speed is low or high, respectively.
It is known that the drag is characterized by the Reynolds number, which is the ratio of the flow speed to the viscosity of the fluid.

On the other hand, the drag in granular environments changes depending on the protocol of the experiments or simulations \cite{Wassgren03,Bharadwaj06,Reddy11,Guillard14,Pal21,Jewel18,Takehara10,Takehara14,Takada17,Takada17_3D,Takada20,Cheng07,Cheng14,Sano12,Sano13,Hilton13,Cruz16,Escalante17,Dhiman20,Hossain20Mobility,Hossain20Rate-dependent,Hossain20Drag,Kubota21} using spheres \cite{Hilton13,Takada17_3D,Takada20,Dhiman20}, cylinders \cite{Wassgren03,Takehara10,Takehara14,Takada17}, or other shapes of objects \cite{Reddy11,Guillard14,Pal21}.
One of the simplest setups was numerically prepared by Hilton and Tordesillas \cite{Hilton13}.
They pulled a spherical intruder in the direction perpendicular to gravity.
In their setup, the drag is given by the combination of a constant term and a term proportional to the drag speed.
Another important setup was studied by Takehara and co-workers \cite{Takehara10,Takehara14}.
They reported that the drag of an intruder in a two-dimensional environment is the sum of a term independent of the drag speed and one depending on the square of the speed.
This was also numerically verified by Takada and Hayakawa in a similar setup \cite{Takada17} where the intruder was pulled in a direction perpendicular to gravity.

We also know that intruders cannot move owing to the existence of surrounding particles if the drag force is sufficiently small.
The threshold is related to the friction between particles or between particles and the bottom \cite{Takada17}.
Although there are few studies on the force below the threshold, it is important to understand this regime because there are fracture and other phenomena.
Once this regime is understood, we can expect to be able to predict the minimum drag force required to move an object, which will yield a reduction of energy loss.
In addition, knowledge of this phenomena may be applied to the application of fracture mechanics or plastic mechanics to granular systems.
Referring to those studies, we previously investigated the drag of the intruder using a similar setup to check how the stress fields around the intruder.
We found good agreement between the stress fields obtained by simulations and those based on the theory of elasticity \cite{Kubota21}.
We note that the drag problems have also been studied in the context of planetary science \cite{Katsuragi}.

Most previous studies focus on the drag of a single intruder.
However, it is important to investigate the drag of multiple intruders \cite{Dhiman20,Escalante17,Cruz16,Espinosa21} when we consider more realistic systems such as the movement of a fork hoe.
In this case, the existence of one intruder must affect the other; therefore, the drag should be different from that in the one-intruder cases.
To clarify this, we study the drag of multiple intruders in a two-dimensional granular environment.
As a first step, we consider two identical cylindrical objects vertically aligned to the drag direction.
In the following, the drag law and the stress around the intruders is investigated by the discrete element method (DEM) \cite{Cundall}.

This paper is organized as follows.
In the next section, we explain the model and the setup of our simulations.
Section \ref{sec:results} is the main part of this paper, in which we show the relationship between the drag force and the steady speed of the intruders.
We also reproduce the stress fields around the intruders from the information on the surface of the intruders.
In Sects.\ \ref{sec:discussion} and \ref{sec:summary}, we discuss and conclude our results.
In Appendix \ref{sec:def_L}, we explain how the system size is determined in the simulations.
In Appendix \ref{sec:three_intruders}, we examine the drag law when we increase the number of intruders.
In Appendix \ref{sec:bipolar}, we present the general framework to obtain the stress fields in terms of the bipolar coordinates.

\section{Model and Setup of Our Simulation}\label{sec:model_setup}
\begin{figure}[htbp]
	\centering
	\includegraphics[width=0.9\linewidth]{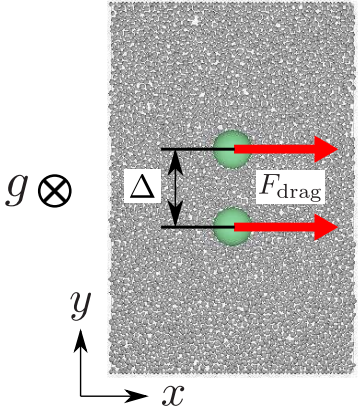}
	\caption{(Color online) Typical snapshot of the simulation.
  	Two intruders are placed with interval $\Delta$.
  	They are pulled with force $F_{\rm drag}$ in the $x$-direction.
	Here, gravity acts perpendicular to the $xy$ plane.}
	\label{fig:setup}
\end{figure}
In this section, we explain the model and the setup of our simulations.
Initially, two identical intruders are placed together at $(x,y)=(0, \pm \Delta/2)$, as shown Fig. \ref{fig:setup}.
Here, the intruders are characterized by diameter $D$ and mass $M$, and the distance between them is fixed.
Then, smaller particles are placed around the intruders at random.
These particles are characterized by diameter $d_i$ and mass $m_i$.
To avoid crystallization, the surrounding particles are prepared with the number ratio $1$$:$$1$ and the size ratio $d$$:$$1.4d$.
We note that the mass ratio is given by $m$:$1.4^2m$.
Here, the mass density of the intruders is assumed to be the same as that of the surrounding particles.
The equations of motion of the intruders and the particles are, respectively, described by \cite{Cundall}
\begin{equation}
  \begin{cases}
    \displaystyle M\frac{d^2\bm{R}_{i}}{dt^2}\cdot\hat{\bm{e}}_x = \sum_j \bm{F}_{ij}\cdot\hat{\bm{e}}_x + F_{\rm drag},\\
    \displaystyle m_i\frac{d^2\bm{r}_i}{dt^2} = \sum_j \bm{F}_{ij} - \mu_{\rm b}m_i g\hat{\bm{v}}_i,
  \end{cases}
\end{equation}
where $\bm{R}_i$ and $\bm{r}_i$ are the position vectors of the $i$-th intruder and the $i$-th particle, respectively, $\bm{F}_{ij}$ is the contact force between the $i$-th and the $j$-th intruders or particles, $F_{\rm{drag}}$ is the drag force acting on the intruders, $\hat{\bm{e}}_x$ is the unit vector parallel to the $x$-direction, $\mu_{\rm b}$ is a coefficient of Coulombic friction acting on the surrounding particles and the bottom plate, $g$ is the acceleration due to gravity, and $\hat{\bm{v}}_i$ is the unit vector parallel to the velocity of the $i$-th particle \cite{Luding08}.
Here, we ignore the motions of the intruders in the $y$-direction.
Hereafter, for simplicity, we do not distinguish $\bm{R}_i$ from $\bm{r}_j$ when we consider the positions of the particles.
The contact force $\bm{F}_{ij}$ is described by the sum of the normal force $F^n_{ij}$ and the tangential force $F^t_{ij}$, i.e., $\bm{F}_{ij} = F^n_{ij}\bm{n} + F^t_{ij}\bm{t}$, where $\bm{n}$ and $\bm{t}$ are unit vectors in the normal and tangential directions between the two contacting particles, respectively.
The force in the normal direction $F^n_{ij}$ is given by
\begin{equation}
  F^n_{ij} = \left[ k_n\delta - \eta_n\left( \bm{v}_{ij}\cdot \hat{\bm{r}}_{ij}\right) \right] \Theta \left( d_{ij}-r_{ij}\right),
\end{equation}
where $k_n$ and $\eta_n$ are the spring constant and the magnitude of the dashpot in the normal direction, respectively, $\delta$ is the overlap of the two contacting particles, $\bm{v}_{ij}$ is the relative velocity described by $\bm{v}_{ij} = \bm{v}_i-\bm{v}_j$, $d_{ij} = (d_i+d_j)/2$ is the sum of the radii of the two contacting particles, $r_{ij} = |\bm{r}_{ij}| = |\bm{r}_i-\bm{r}_j|$ is the relative displacement, $\hat{\bm{r}}_{ij} = \bm{r}_{ij}/r_{ij}$, and $\Theta(x)$ is the step function.
Here, it is well known that the viscoelastic force becomes linear with a logarithmic correction, which means that the linear model is better than the Hertzian model \cite{Kuninaka01, Hayakawa02}.
Next, the force in the tangential direction $F^t_{ij}$ is given by
\begin{equation}
  F^t_{ij} = \min \left( \mu F^n_{ij}, \left|-k_t\bm{\xi} -\eta_t\bm{v}^t_{ij}\right| \right)\Theta \left( d_{ij}-r_{ij}\right),
\end{equation}
where $\mu$ is the coefficient of Coulombic friction acting between the particles, $k_t$ and $\eta_t$ are the spring constant and the magnitude of the dashpot in the tangential direction, and $\bm{v}^t_{ij} = \bm{v}_{ij}-(\bm{v}_{ij}\cdot \hat{\bm{r}}_{ij})\hat{\bm{r}}_{ij}$ and $\bm{\xi} = \int^t\bm{v}^t_{ij}(t^{\prime})dt^{\prime}$ are the tangential relative velocity and the tangential displacement from the beginning of the contact, respectively \cite{Luding08}.
In the simulation, we set $D = 10d$, number of the surrounding particles $N = 4000$, packing fraction $\varphi = 0.80$, $k_t = k_n$, $\eta_t = \eta_n$, $\mu = \mu_{\rm b} = 0.20$, and $g = 1.0 \times 10^{-4} k_nd/m$.
For our choice of parameters, the restitution coefficients in the normal and tangential directions correspond to $e_n=e_t=4.3\times10^{-2}$.
Here, the determination of the system sizes ($L_x$ and $L_y$) is discussed in Appendix \ref{sec:def_L}.
We apply the periodic boundary condition in the $x$-direction, and we put the smaller particles at $y = \pm L_y / 2$ as the bumpy boundary condition in the $y$-direction \cite{Takada17}.
In this paper, we mainly focus on the case where there exist two intruders in the system.
The results for three intruders are briefly discussed in Appendix \ref{sec:three_intruders}.

\section{Results}\label{sec:results}
In this section, we explain the results obtained from the simulation in two parts.
In the first subsection, we show the drag law when the drag force is sufficiently larger than the yield force, which will be defined there.
In the next subsection, we focus on the stress fields when the drag force is smaller than but close to the yield force.
We try to reproduce the stress fields around the intruders in terms of the Airy stress function known in elasticity.

\begin{figure}
	\centering
	\includegraphics[width=\linewidth]{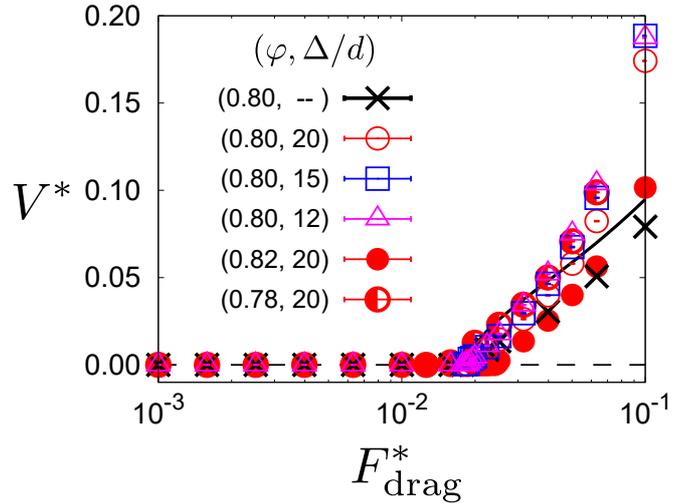}
	\caption{(Color online) Steady speed of the intruders against the drag force for various distances $\Delta$ and the packing fraction $\varphi$ for $D=10d$.
  Here, we have introduced $F_{\rm drag}^* \equiv F_{\rm drag}/(k_n d)$ and $V^*\equiv V/(d \sqrt{k_n/m})$.
  The solid line represents Eq.\ \eqref{eq:FV} for the single-intruder case.}
	\label{fig:FV}
\end{figure}

\subsection{Drag law}\label{sec:drag_law}
First, we observe the steady speed $V$ when we change the drag force $F_{\rm drag}$, the distance of the intruders $\Delta$, and the packing fraction $\varphi$.
Note that the rotational motion of the intruders is not observed.
As shown in Fig.\ \ref{fig:FV}, there is a yield force $F_{\rm Y}$, and the value of $F_{\rm Y}$ is found to be almost unchanged for the distance $\Delta$, although it increases as the packing fraction $\varphi$ becomes larger. 
We note that the velocity dependence is almost the same. 
Now, we focus on the dependence of the drag on the distance $\Delta$.

\begin{figure}
	\centering
	\includegraphics[width=\linewidth]{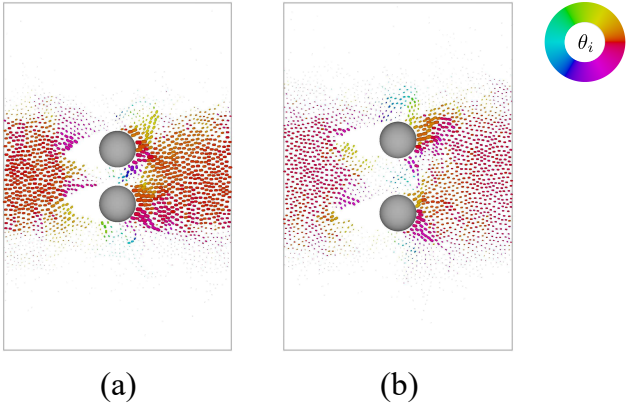}
	\caption{(Color online) Displacements of particles with the time interval $\Delta t = 50\sqrt{m/k_n}$ in the steady state for $\varphi = 0.80$, $D = 10d$, and $F_{\rm drag} = 2.5 \times 10^{-2} k_nd$. 
  Panels (a) and (b) correspond to $\Delta = 15d$ and $20d$, respectively. 
  The color represents the angle of the displacements.
  The line width is propotional to the displacement.}
	\label{fig:displacement}
\end{figure}

For $F_{\rm drag} < F_{\rm Y}$, the intruders cannot move because the drag force and the frictional forces of the surrounding particles are balanced.
From the simulation results, the value of the yield force becomes $F_{\rm Y}\simeq 1.8\times10^{-2}k_nd$ for $\mu_{\rm b} = 0.20$.
Above the yield force $F_{\rm Y}$, on the other hand, the intruders can move with a steady speed $V$, where $V$ is an increasing function depending on $F_{\rm drag}$.
Note that the rate of increase of $V$ depends on the distance $\Delta$.
Here, in the case of a single intruder, we previously reported that the relationship between the drag force $F_{\rm drag}$ and the steady speed $V$ is given by \cite{Kubota21}
\begin{equation}
  F_{\rm drag} = F_{\rm Y} + \frac{2}{\pi}\varphi\frac{Mm}{M+m}\frac{D+d}{d^2}V^2\left(\frac{14}{9}+\frac{4}{3}e_n+\frac{2}{9}e_t\right),
  \label{eq:FV}
\end{equation}
which is derived under the approximation that the surrounding particles are monodisperse and the collisions are characterized by the restitution coefficients $e_n$ and $e_t$ (see, e.g., Ref.\ \citenum{Garzo_book}).
When there exist two intruders, the steady speed becomes higher as the interval between the intruders decreases, which was also reported in Ref.\ \citenum{Dhiman20}.
This is because the area pushed by the upper intruder has a part that is common with that of the lower intruder.
This fact is also understood from Fig.\ \ref{fig:displacement}, which shows the displacements of particles with the time interval $\Delta t=50\sqrt{m/k_n}$ when we change the interval between the intruders, $\Delta$.
It is clear that the particles between the intruders are pushed by both intruders.
Note that the drag might change when the distance between the intruders becomes smaller as $\Delta-D\gtrsim d$ because we keep its distance unchanged.
For the sufficiently small interval, the surrounding particles cannot pass between the intruders \cite{Dhiman20}.

\begin{figure}
	\centering
	\includegraphics[width=\linewidth]{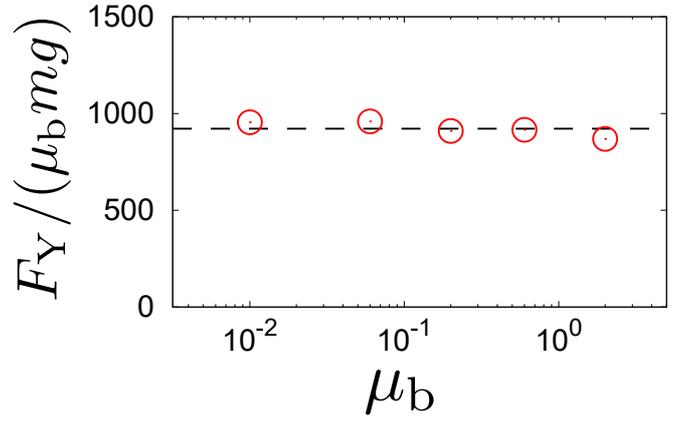}
	\caption{(Color online) Yield force of the intruders against frictional coefficient $\mu_{\rm b}$.
  The dashed line represents $F_{\rm Y} = 9.2\times 10^{2}\mu_{\rm b}mg$.}
	\label{fig:mu_fy}
\end{figure}
Next, we observe the yield force $F_{\rm Y}$ of the intruders for various bottom friction coefficients $\mu_{\rm b}$ when we fix $\Delta = 15d$.
The yield force $F_{\rm Y}$ is related to the frictional forces of the surrounding particles interacting with the intruders.
This means that the yield force should be given by
\begin{equation}
    F_{\rm Y} = \mu_{\rm b}M_{\rm Y} g,
\end{equation}
where $M_{\rm Y}$ is the total mass of all particles interacting with the intruders.
Figure \ref{fig:mu_fy} indicates that the above interpretation is correct because the ratio of the yield force to the frictional force is almost constant.

\subsection{Stress fields around intruders}
For $F_{\rm drag}<F_{\rm Y}$, let us investigate the stress fields around the intruders.
Here, the stress obtained from the simulation is given by
\begin{equation}
  \sigma_{\alpha\beta} = \frac{1}{L_xL_y}\sum_i\left(m_iv_{i,\alpha}v_{i,\beta}
  + \frac{1}{2}\sum_j r_{ij,\alpha}F_{ij,\beta}\right),
\end{equation}
where we have used the coarse-grained method for smoothing the data with the characteristic length $w=2d$\cite{Zhang10}.

\begin{figure}
	\centering
	\includegraphics[width=\linewidth]{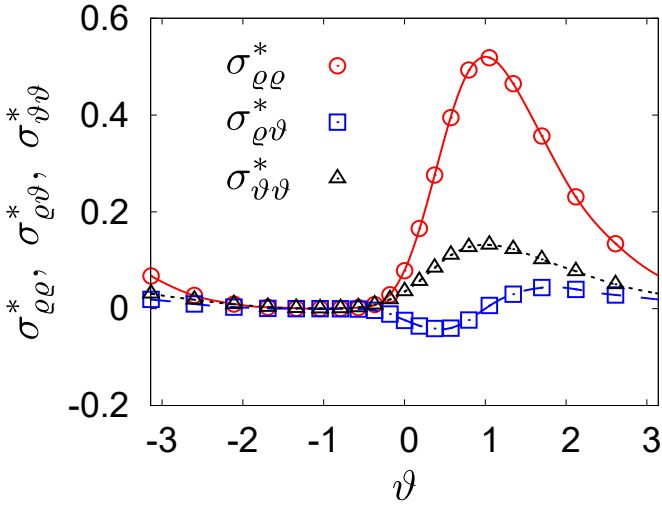}
	\caption{(Color online) Profiles of the stresses on the surfaces of the upper intruders for $D = 10d$, $\Delta = 20d$, and $F_{\rm drag} = 1.6 \times 10^{-2}k_n d$, where we have introduced $\sigma_{\alpha\beta}^*=\sigma_{\alpha\beta}/(F_{\rm drag}/(\pi d))$ with $(\alpha,\beta)=(\varrho,\varrho)$, $(\varrho,\vartheta)$, and $(\vartheta,\vartheta)$.
  The marks represent the stresses obtained from the simulations, and the lines represent the Fourier series expansions of the stresses.}
	\label{fig:stress of surface}
\end{figure}

We introduce the Airy stress function for comparison with the theory of elasticity \cite{Timoshenko,Daniels17}.
For this purpose, let us expand $\sigma_{\varrho\varrho}$, $\sigma_{\varrho\vartheta}$, and $\sigma_{\vartheta\vartheta}$ in the Fourier series as
\begin{equation}
  \begin{Bmatrix}
    \sigma_{\varrho\varrho}\\
    \sigma_{\varrho\vartheta}\\
    \sigma_{\vartheta\vartheta}
  \end{Bmatrix}
  =\frac{1}{2}
  \begin{Bmatrix}
    \widetilde{\sigma}_{\varrho\varrho}^{{\rm C}(0)}\\
    \widetilde{\sigma}_{\varrho\vartheta}^{{\rm C}(0)}\\
    \widetilde{\sigma}_{\vartheta\vartheta}^{{\rm C}(0)}
  \end{Bmatrix}
  + \sum_{n=1}^8
  \left[
    \begin{Bmatrix}
      \widetilde{\sigma}_{\varrho\varrho}^{{\rm C}(n)}\\
      \widetilde{\sigma}_{\varrho\vartheta}^{{\rm C}(n)}\\
      \widetilde{\sigma}_{\vartheta\vartheta}^{{\rm C}(n)}
    \end{Bmatrix}
    \cos(n\vartheta)
    +
    \begin{Bmatrix}
      \widetilde{\sigma}_{\varrho\varrho}^{{\rm S}(n)}\\
      \widetilde{\sigma}_{\varrho\vartheta}^{{\rm S}(n)}\\
      \widetilde{\sigma}_{\vartheta\vartheta}^{{\rm S}(n)}
    \end{Bmatrix}
    \sin(n\vartheta)
  \right].
\end{equation}
In the Fourier series, we consider terms up to the eighth order.
As shown in Fig.\ \ref{fig:stress of surface}, the Fourier series up to the eighth order well reproduce those obtained from the simulations without any fitting parameter.
We mainly focus on the upper intruder throughout this paper because the stress fields are symmetric with respect to the $x$-axis.

Now, let us construct the stress field around the intruders in terms of the surface stress.
As discussed in Appendix \ref{sec:bipolar}, it is convenient to adopt the bipolar coordinates as
\begin{equation}
\begin{cases}
	\displaystyle x = \frac{\sqrt{\Delta^2-D^2}}{2}\frac{\sin\vartheta}{\cosh\varrho - \cos\vartheta}\\
	\displaystyle y = \frac{\sqrt{\Delta^2-D^2}}{2}\frac{\sinh\varrho}{\cosh\varrho - \cos\vartheta}
\end{cases}.
	\label{eq:xy_alphabeta}
\end{equation}
Using the same procedure as in Ref.\ \citenum{Kubota21}, the stress field can be constructed in terms of the Airy stress function $\chi$ as \cite{Jeffery21}
\begin{subequations}\label{eq:sigma_aa, bb, ab}
\begin{align}
	\sigma_{\varrho\varrho}
	&= \frac{2}{\sqrt{\Delta^2-D^2}}
  \left[(\cosh\varrho-\cos\vartheta)\frac{\partial^2}{\partial \vartheta^2} 
		-\sinh\varrho \frac{\partial}{\partial \varrho} \right.\nonumber\\
    &\hspace{6em}\left.-\sin\vartheta\frac{\partial}{\partial \vartheta}
		+\cosh\varrho\right]\left(\frac{\chi}{J}\right)\\
	\sigma_{\vartheta\vartheta}
	&= \frac{2}{\sqrt{\Delta^2-D^2}}
  \left[(\cosh\varrho-\cos\vartheta)\frac{\partial^2}{\partial \varrho^2} 
		-\sinh\varrho \frac{\partial}{\partial \varrho} \right.\nonumber\\
    &\hspace{6em}\left.-\sin\vartheta\frac{\partial}{\partial \vartheta}
		+\cos \vartheta\right]\left(\frac{\chi}{J}\right)\\
	\sigma_{\varrho\vartheta}
	&= -\frac{2}{\sqrt{\Delta^2-D^2}}
  (\cosh\varrho-\cos\vartheta)\frac{\partial^2}{\partial \varrho \partial \vartheta}\left(\frac{\chi}{J}\right),
\end{align}
\end{subequations}
where
\begin{equation}\label{eq:J}
	J = \frac{\sqrt{\Delta^2-D^2}}{2(\cosh\varrho-\cos\vartheta)},
\end{equation}
and
\begin{align}
	\frac{\chi}{J}
	&= (E\cos\vartheta + F \sin\vartheta + G\cosh\varrho + H\sinh\varrho)\vartheta\nonumber\\
		&\hspace{1em}
		+ (B_0\cosh\varrho + D_0\sinh\varrho)\varrho\nonumber\\
		&\hspace{1em}
		+ [A_1 \cosh(2\varrho) + B_1 + C_1 \sinh(2\varrho) + D_1\varrho]\cos\vartheta\nonumber\\
		&\hspace{1em}
		+ [A_1^\prime \cosh(2\varrho) + C_1^\prime \sinh(2\varrho) + D_1^\prime\varrho]\sin\vartheta\nonumber\\
		&\hspace{1em}
		+ \sum_{n=2}^8
			\left\{A_n \cosh[(n+1)\varrho] + B_n \cosh[(n-1)\varrho]\right.\nonumber\\
		&\hspace{2em}\left.
				+C_n \sinh[(n+1)\varrho] + D_n\sinh[(n-1)\varrho]\right\}\cos(n\vartheta)\nonumber\\
		&\hspace{1em}
		+ \sum_{n=2}^8
			\left\{A_n^\prime \cosh[(n+1)\varrho] + B_n^\prime \cosh[(n-1)\varrho]\right.\nonumber\\
		&\hspace{2em}\left.
				+C_n^\prime \sinh[(n+1)\varrho] + D_n^\prime\sinh[(n-1)\varrho]\right\}\sin(n\vartheta),
        \label{eq:chi}
\end{align}
with constants $E$, $F$, $G$, $H$, $A_n$, $B_n$, $C_n$, $D_n$, $A_n^\prime$, $B_n^\prime$, $C_n^\prime$, and $D_n^\prime$ ($n\ge0$).
Note that higher order terms ($n\ge9$) are dropped in Eq.\ \eqref{eq:chi} because they do not affect the results.
Here, we consider the functional form of each stress component, $H$, $B_0$, $B_0^\prime$ $C_n$, $D_n$, $C_n^\prime$, and $D_n^\prime$ to be zero.

\begin{figure}
	\centering
	\includegraphics[width=\linewidth]{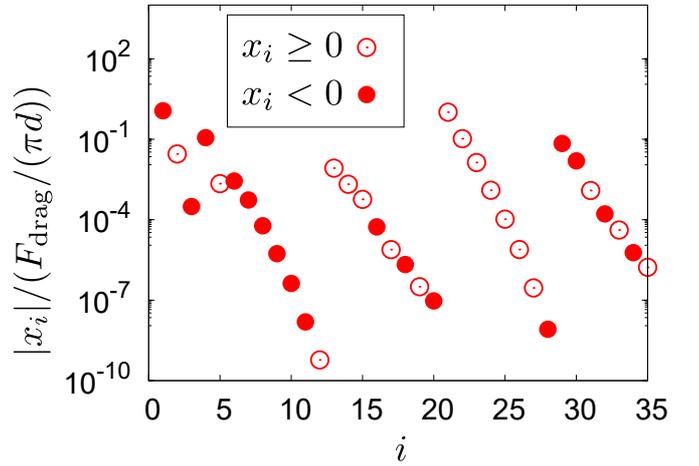}
	\caption{(Color online) Magnitude of each component $|x_i|$ ($i=1,\cdots, 35$) for $\varphi=0.80$, $D=10d$, $\Delta=20d$, and $F_{\rm drag}=1.6\times 10^{-2}k_n d$.
  Here, the open (filled) circles represent positive (negative) $x_i$.}
	\label{fig:coeff}
\end{figure}

To determine each Fourier coefficient, we introduce $35$-dimensional vectors $\bm{\widetilde{\Sigma}}_{\rm{surf}}=(\widetilde{\sigma}_{\varrho\varrho}^{{\rm C}(0)},\ \cdots,\ \widetilde{\sigma}_{\varrho\varrho}^{{\rm C}(8)},\ \widetilde{\sigma}_{\varrho\varrho}^{{\rm S}(1)},$
$\cdots,\ \widetilde{\sigma}_{\varrho\varrho}^{{\rm S}(8)},\ \widetilde{\sigma}_{\varrho\vartheta}^{{\rm C}(0)},\ \cdots,\ \widetilde{\sigma}_{\varrho\vartheta}^{{\rm C}(8)},\ \widetilde{\sigma}_{\varrho\vartheta}^{{\rm S}(1)},\ \cdots,\ \widetilde{\sigma}_{\varrho\vartheta}^{{\rm S}(8)},\ \widetilde{\sigma}_{\varrho\varrho}^{{\rm C}(0)}+\widetilde{\sigma}_{\vartheta\vartheta}^{{\rm C}(0)})^T$ and $\bm{X}=(x_1, x_2, \cdots, x_{35})^T$.
Here, the components of $\bm{X}$ are given by
\begin{align}
  &x_1 = E.\quad
  x_2 = F,\quad
  x_3 = G,\quad
  x_4 = D_0,\nonumber\\
  &x_{5} = A_1,\ \cdots,\ x_{12} = A_8,\ 
  x_{13} = A^\prime_1,\ \cdots,\ x_{20} = A^\prime_8,\nonumber\\
  &x_{21} = B_1,\ \cdots,\ x_{28} = B_8,\ 
  x_{29} = B^\prime_2,\ \cdots,\ x_{35} = B^\prime_8.
\end{align}
Here, the Fourier components are functions of the vector $\bm{X}$, where their expressions are summarized in Table \ref{fig:coeff_table}.
\begin{table*}[htbp]
	\caption{Coefficients in the Fourier series of $\sigma_{\varrho\varrho}$, $\sigma_{\varrho\vartheta}$, and $\sigma_{\vartheta\vartheta}$ in Eqs.\ \eqref{eq:sigma_aa, bb, ab} with Eq.\ \eqref{eq:chi}.} 
	\label{fig:coeff_table}
	\begin{tabular}{l}
	\hline
	{$a_{\varrho\varrho}^{(0)}
	= - 3F +D_0 + 2B_1 - (D_0-2A_1) \cosh (2\varrho)$}\quad
	{$a_{\varrho\varrho}^{(1)}
	= (2F +A_1 + 3B_2) \cosh(\varrho) 
	- (A_1- 3A_2) \cosh(3\varrho)$}\\
	{$a_{\varrho\varrho}^{(2)}
	= -\frac{1}{2}F - B_2 - 2(B_2 - 3B_3)\cosh(2\varrho) 
	- 3(A_2 - 2A_3) \cosh(4 \varrho)$}\quad
	{$a_{\varrho\varrho}^{(3)}
	= (B_2 - 3B_3)\cosh(\varrho) 
	+ (A_2 - 2A_3 - 5B_3 + 10B_4)\cosh(3\varrho) 
	- 2(3A_3 - 5A_4)\cosh(5\varrho)$}\\
	{$a_{\varrho\varrho}^{(4)}
	= 3(B_3 - 2B_4)\cosh(2\varrho) 
	+ (3A_3 - 5A_4 - 9B_4 + 15B_5)\cosh(4\varrho) 
	- 5(2A_4 - 3A_5)\cosh(6\varrho)$}\\
	{$a_{\varrho\varrho}^{(5)}
	= 2(3B_4 - 5B_5)\cosh(3\varrho) 
	+ (6A_4 - 9A_5 - 14B_5 + 21B_6)\cosh(5\varrho) 
	- 3(5A_5 - 7A_6) \cosh(7\varrho)$}\\
	{$a_{\varrho\varrho}^{(6)}
	= 5(2B_5 - 3B_6)\cosh(4\varrho) 
	+ 2(5A_5 - 7A_6 - 10B_6 + 14B_7)\cosh(6\varrho) 
	- 7(3A_6 - 4A_7)\cosh(8\varrho)$}\\
	{$a_{\varrho\varrho}^{(7)}
	= 3(5B_6 - 7B_7)\cosh(5\varrho) 
	+ (15A_6 - 20A_7 - 27B_7 + 36B_8)\cosh(7\varrho) 
	- 4(7A_7 - 9A_8)\cosh(9\varrho)$}\\
	{$a_{\varrho\varrho}^{(8)}
	= 7(3B_7 - 4B_8)\cosh(6\varrho) 
	+ (21A_7 - 27A_8 - 35B_8)\cosh(8\varrho) 
	- 36A_8\cosh(10\varrho)$}\\ \hline
	{$b_{\varrho\varrho}^{(1)}
	= 2(E + G) 
	+ (-2E - G + A^\prime_1 + 3B^\prime_2)\cosh(\varrho) 
	- (A^\prime_1 - 3A^\prime_2)\cosh(3\varrho)$}\quad
	{$b_{\varrho\varrho}^{(2)}
	= -\frac{1}{2}E - G - B^\prime_2 
	- 2(B^\prime_2 - 3B^\prime_3)\cosh(2\varrho) 
	- 3(A^\prime_2 - 2A^\prime_3)\cosh(4\varrho)$}\\
	{$b_{\varrho\varrho}^{(3)}
	= \frac{2}{3}(E + G) 
	+ (B^\prime_2 - 3B^\prime_3)\cosh(\varrho) 
	+ (A^\prime_2 - 2A^\prime_3 - 5B^\prime_3 + 10B^\prime_4)\cosh(3\varrho) 
	- 2(3A^\prime_3 - 5A^\prime_4)\cosh(5\varrho)$}\\
	{$b_{\varrho\varrho}^{(4)}
	= -\frac{1}{2}(E + G) 
	+ 3(B^\prime_3 - 2B^\prime_4)\cosh(2\varrho) 
	+ (3A^\prime_3 - 5A^\prime_4 - 9B^\prime_4 + 15 B^\prime_5)\cosh(4\varrho) 
	- 5 (2A^\prime_4 - 3A^\prime_5)\cosh(6\varrho)$}\\
	{$b_{\varrho\varrho}^{(5)}
	= \frac{2}{5}(E + G) 
	+ 2(3B^\prime_4 - 5B^\prime_5)\cosh(3\varrho) 
	+ (6A^\prime_4 - 9A^\prime_5 - 14B^\prime_5 + 21B^\prime_6)\cosh(5\varrho) 
	- 3(5A^\prime_5- 7A^\prime_6)\cosh(7\varrho)$}\\
	{$b_{\varrho\varrho}^{(6)}
	= -\frac{1}{3} (E+G)
	+ 5(2B^\prime_5 - 3B^\prime_6) \cosh (4 \varrho)
	+2 (5 A^\prime_5-7 A^\prime_6-10 B^\prime_6+14 B^\prime_7) \cosh (6 \varrho )
	-7(3 A^\prime_6 - 4 A^\prime_7) \cosh (8 \varrho)$}\\
	{$b_{\varrho\varrho}^{(7)}
	= \frac{2}{7} (E+G) 
	+ 3 (5 B^\prime_6-7 B^\prime_7) \cosh (5 \varrho )
	+ (15 A^\prime_6-20 A^\prime_7-27 B^\prime_7+36B^\prime_8) \cosh (7\varrho) 
	- 4 (7 A^\prime_7-9 A^\prime_8) \cosh (9 \varrho)$}\\
	{$b_{\varrho\varrho}^{(8)}
	= -\frac{1}{4}(E + G) 
	+ 7 (3 B^\prime_7-4 B^\prime_8) \cosh (6 \varrho) 
	+ (21 A^\prime_7-27 A^\prime_8-35 B^\prime_8)\cosh (8 \varrho)
	- 36 A^\prime_8 \cosh (10 \varrho)$}\\ \hline
	{$a_{\varrho\vartheta}^{(0)}
	= -(G - 2A^\prime_1)\sinh(2\varrho)$}\quad
	{$a_{\varrho\vartheta}^{(1)}
	= (G -A^\prime_1 + B^\prime_2)\sinh(\varrho) 
	- (A^\prime_1 - 3A^\prime_2)\sinh(3\varrho)$}\quad
	{$a_{\varrho\vartheta}^{(2)}
	= (A^\prime_1 - 3A^\prime_2 - B^\prime_2 + 3B^\prime_3)\sinh(2\varrho) 
	- 3(A^\prime_2 - 2A^\prime_3)\sinh(4\varrho)$}\\
	{$a_{\varrho\vartheta}^{(3)}
	= (B^\prime_2 - 3B^\prime_3)\sinh(\varrho) 
	+ 3(A^\prime_2 - 2A^\prime_3 -B^\prime_3 + 2B^\prime_4)\sinh(3\varrho) 
	- 2(3A^\prime_3 - 5A^\prime_4)\sinh(5\varrho)$}\\
	{$a_{\varrho\vartheta}^{(4)}
	= 3(B^\prime_3 - 2B^\prime_4) \sinh(2\varrho)
	+2 (3A^\prime_3 - 5A^\prime_4 - 3B^\prime_4 + 5B^\prime_5) \sinh(4\varrho) 
	- 5(2A^\prime_4 - 3A^\prime_5)\sinh(6\varrho)$}\\
	{$a_{\varrho\vartheta}^{(5)}
	= 2(3B^\prime_4 - 5B^\prime_5) \sinh(3\varrho)
	+ 5(2A^\prime_4 - 3 A^\prime_5 - 2 B^\prime_5 + 3B^\prime_6) \sinh(5\varrho) 
	- 3(5A^\prime_5 -7A^\prime_6)\sinh(7\varrho)$}\\
	{$a_{\varrho\vartheta}^{(6)}
	= 5(2B^\prime_5 - 3B^\prime_6) \sinh(4\varrho)
	+ 3(5A^\prime_5 - 7A^\prime_6 - 5B^\prime_6 + 7B^\prime_7)\sinh(6\varrho)
	- 7(3A^\prime_6 - 4A^\prime_7)\sinh(8\varrho)$}\\
	{$a_{\varrho\vartheta}^{(7)}
	= 3(5B^\prime_6 - 7B^\prime_7) \sinh(5\varrho)
	+ 7 (3A^\prime_6 - 4A^\prime_7 - 3B^\prime_7 + 4B^\prime_8) \sinh(7\varrho)
	- 4(7A^\prime_7 - 9A^\prime_8)\sinh(9\varrho)$}\\
	{$a_{\varrho\vartheta}^{(8)}
	= 7(3B^\prime_7 - 4B^\prime_8) \sinh(6\varrho)
	+ 4(7A^\prime_7 - 9A^\prime_8 - 7B^\prime_8) \sinh(8\varrho)
	- 36A^\prime_8\sinh(10\varrho)$}\\ \hline
	{$b_{\varrho\vartheta}^{(1)}
	= (A_1 - B_2)\sinh(\varrho) 
	+ (A_1 - 3A_2)\sinh(3\varrho)$}\quad
	{$b_{\varrho\vartheta}^{(2)}
	= -(A_1 - 3A_2 - B_2 + 3B_3)\sinh(2\varrho) 
	+ 3(A_2 - 2A_3)\sinh(4\varrho)$}\\
	{$b_{\varrho\vartheta}^{(3)}
	= -(B_2 - 3B_3)\sinh(\varrho) 
	- 3(A_2 - 2A_3 - B_3 + 2B_4)\sinh(3\varrho) 
	+ 2(3A_3 - 5A_4)\sinh(5\varrho)$}\\
	{$b_{\varrho\vartheta}^{(4)}
	= -3(B_3 - 2B_4)\sinh(2\varrho)
	- 2(3A_3 - 5A_4 - 3B_4 + 5B_5)\sinh(4\varrho) 
	+ 5(2A_4 - 3A_5)\sinh(6\varrho)$}\\
	{$b_{\varrho\vartheta}^{(5)}
	= -2(3B_4 - 5B_5)\sinh(3\varrho) 
	- 5(2A_4 - 3A_5 - 2B_5 + 3B_6)\sinh(5\varrho) 
	+ 3(5A_5 - 7A_6)\sinh(7\varrho)$}\\
	{$b_{\varrho\vartheta}^{(6)}
	= -5(2B_5 - 3B_6)\sinh(4\varrho) 
	- 3(5A_5 - 7A_6 - 5B_6 + 7B_7)\sinh(6\varrho) 
	+ 7(3A_6 - 4A_7)\sinh(8\varrho)$}\\
	{$b_{\varrho\vartheta}^{(7)}
	= -3(5B_6 - 7B_7)\sinh(5\varrho) 
	- 7(3A_6 - 4A_7 - 3B_7 + 4B_8)\sinh(7\varrho) 
	+ 4(7A_7 - 9A_8)\sinh(9\varrho)$}\\
	{$b_{\varrho\vartheta}^{(8)}
	= -7(3B_7 - 4B_8)\sinh(6\varrho) 
	- 4(7A_7 - 9A_8 - 7B_8)\sinh(8\varrho) 
	+ 36A_8\sinh(10\varrho)$}\\ \hline
	{$a_{\vartheta\vartheta}^{(0)}
	= -F + 2B_1 + 3D_0 
	+ (D_0 - 2A_1)\cosh(2\varrho)$}\quad
	{$a_{\vartheta\vartheta}^{(1)}
	= (-2D_0 + 3A_1 + B_2)\cosh(\varrho) 
  	+ (A_1 - 3A_2)\cosh(3\varrho)$}\\
	{$a_{\vartheta\vartheta}^{(2)}
	= \frac{1}{2}(F + 2B_2) 
	- 2(A_1 - 3A_2)\cosh(2\varrho) 
	+ 3(A_2 - 2A_3)\cosh(4\varrho)$}\quad
	{$a_{\vartheta\vartheta}^{(3)}
	= -(B_2 - 3B_3)\cosh(\varrho) 
	- (5A_2 - 10A_3 - B_3 + 2B_4)\cosh(3\varrho) 
	+ 2(3A_3 - 5A_4)\cosh(5\varrho)$}\\
	{$a_{\vartheta\vartheta}^{(4)}
	= -3(B_3 - 2B_4)\cosh(2\varrho) 
	- (9A_3 - 15A_4 - 3B_4 + 5B_5)\cosh(4\varrho) 
	+ 5(2A_4 - 3A_5)\cosh(6\varrho)$}\\
	{$a_{\vartheta\vartheta}^{(5)}
	= -2(3B_4 - 5B_5)\cosh(3\varrho) 
	- (14A_4 - 21A_5 - 6B_5 + 9B_6)\cosh(5\varrho) 
	+ 3(5A_5 - 7A_6)\cosh(7\varrho)$}\\
	{$a_{\vartheta\vartheta}^{(6)}
	= -5(2B_5 - 3B_6)\cosh(4\varrho) 
	- 2(10A_5 - 14A_6 - 5B_6 + 7B_7)\cosh(6\varrho) 
	+ 7(3A_6 - 4A_7)\cosh(8\varrho)$}\\
	{$a_{\vartheta\vartheta}^{(7)}
	= -3(5B_6 - 7B_7)\cosh(5\varrho) 
	- (27A_6 - 36A_7 - 15B_7 + 20B_8)\cosh(7\varrho) 
	+ 4(7A_7 - 9A_8)\cosh(9\varrho)$}\\
	{$a_{\vartheta\vartheta}^{(8)}
	= -7(3B_7 - 4B_8)\cosh(6\varrho) 
	- (35A_7 - 45A_8 - 21B_8)\cosh(8\varrho) 
	+ 36A_8\cosh(10\varrho)$}\\ \hline
	{$b_{\vartheta\vartheta}^{(1)}
	= 2(E + G) 
	- (G - 3A^\prime_1 - B^\prime_2)\cosh(\varrho) 
	+ (A^\prime_1 - 3A^\prime_2)\cosh(3\varrho)$}\quad
	{$b_{\vartheta\vartheta}^{(2)}
	= -\frac{1}{2}(3E + 2G - 2B^\prime_2) 
	- 2(A^\prime_1 - 3A^\prime_2)\cosh(2\varrho) 
	+ 3(A^\prime_2 - 2A^\prime_3)\cosh(2\varrho)$}\\
	{$b_{\vartheta\vartheta}^{(3)}
	= \frac{2}{3}(E + G)
	- (B^\prime_2 - 3B^\prime_3)\cosh(\varrho) 
	- (5A^\prime_2 - 10A^\prime_3 - B^\prime_3 + 2B^\prime_4)\cosh(3\varrho)
	+ 2(3A^\prime_3 - 5A^\prime_4)\cosh(5\varrho)$}\\
	{$b_{\vartheta\vartheta}^{(4)}
	= -\frac{1}{2}(E + G)
	- 3(B^\prime_3 - 2B^\prime_4)\cosh(2\varrho) 
	- (9A^\prime_3 - 15A^\prime_4 - 3B^\prime_4 + 5B^\prime_5)\cosh(4\varrho)
	+ 5(2A^\prime_4 - 3A^\prime_5)\cosh(6\varrho)$}\\
	{$b_{\vartheta\vartheta}^{(5)}
	= \frac{2}{5}(E + G)
	- 2(3B^\prime_4 - 5B^\prime_5)\cosh(3\varrho) 
	- (14A^\prime_4 - 21A^\prime_5 - 6B^\prime_5 + 9B^\prime_6)\cosh(5\varrho)
	+ 3(5A^\prime_5 - 7A^\prime_6)\cosh(7\varrho)$}\\
	{$b_{\vartheta\vartheta}^{(6)}
	= -\frac{1}{3}(E + G)
	- 5(2B^\prime_5 - 3B^\prime_6)\cosh(4\varrho) 
	- 2(10A^\prime_5 - 14A^\prime_6 - 5B^\prime_6 + 7B^\prime_7)\cosh(6\varrho)
	+ 7(3A^\prime_6 - 4A^\prime_7)\cosh(8\varrho)$}\\
	{$b_{\vartheta\vartheta}^{(7)}
	= \frac{2}{7}(E + G)
	- 3(5B^\prime_6 - 7B^\prime_7)\cosh(5\varrho) 
	- (27A^\prime_6 - 36A^\prime_7 - 15B^\prime_7 + 20B^\prime_8)\cosh(7\varrho)
	+ 4(7A^\prime_7 - 9A^\prime_8)\cosh(9\varrho)$}\\
	{$b_{\vartheta\vartheta}^{(8)}
	= -\frac{1}{4}(E + G)
	- 7(3B^\prime_7 - 4B^\prime_8)\cosh(6\varrho) 
	- (35A^\prime_7 - 45A^\prime_8 - 21B^\prime_8)\cosh(8\varrho)
	+ 36A^\prime_8\cosh(10\varrho)$}\\ \hline
	\end{tabular}
\end{table*}
When we introduce a coefficient matrix $C$, we can write simultaneous equations to determine the vector $\bm{X}$ as
\begin{equation}
  C \bm{X} = \widetilde{\bm{\Sigma}}_{\rm surf}.
\end{equation}
This means that we can evaluate $\bm{X}=C^{-1}\widetilde{\bm{\Sigma}}_{\rm surf}$ from the information of the surfaces of the intruders.
Figure \ref{fig:coeff} shows the absolute value of each component $|x_i|$.
This numerically justifies the truncation of higher order terms in Eq.\ \eqref{eq:chi} because the values become exponentially smaller for larger order.
Figure \ref{fig:coeff} shows that only $x_1 = E$ and $x_{21} = B_1$ have large absolute values, while the rest are much smaller and almost zero.
This means that $E$ and $B_1$ play dominant roles in the stress fields.
We will revisit this in the next section.

\begin{figure}[htbp]
	\centering
	\includegraphics[width=0.85\linewidth]{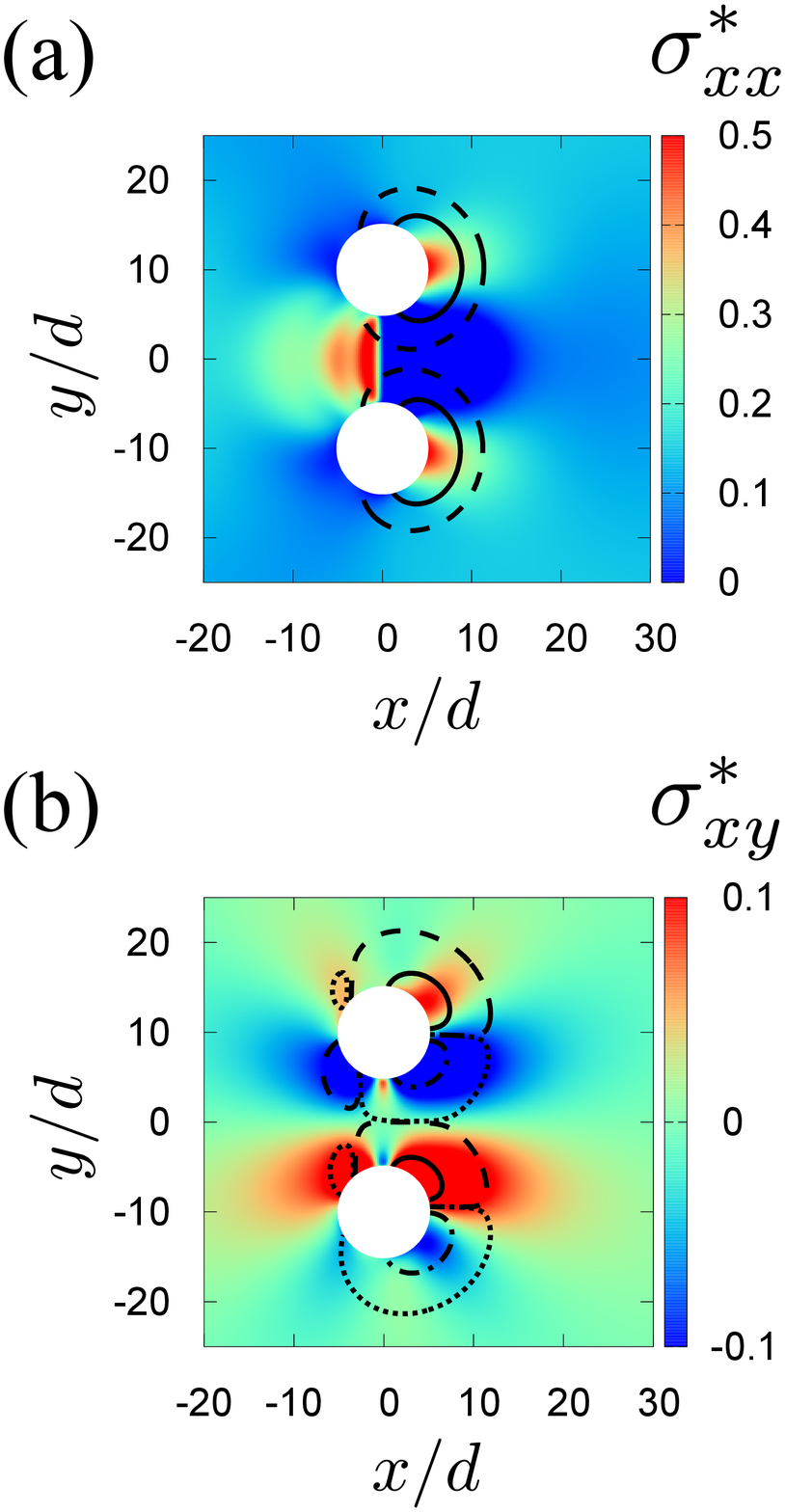}
	\caption{(Color online) Spatial distributions of (a) $\sigma_{xx}$ and (b) $\sigma_{xy}$ obtained from the simulations and from the Airy stress function \eqref{eq:chi} with the coefficients in Fig.\ \ref{fig:coeff} for $\varphi=0.80$, $D=10d$, $\Delta=20d$, and $F_{\rm drag}=1.6\times 10^{-2}k_n d$.
	The solid and dashed lines in panel (a) represent the contours of $5.0\times 10^{-2}$ and $2.0\times10^{-3}$, respectively.
	The solid, dashed, dotted, and dot--dashed lines in panel (b) correspond to the contours of $3.0\times10^{-2}$, $1.0\times 10^{-4}$, $-1.0\times 10^{-4}$, and $-3.0\times10^{-2}$, respectively.}
	\label{fig:stress_fields}
\end{figure}

Now, let us check whether the above estimation is valid.
Figure \ref{fig:stress_fields} shows the stress fields $\sigma_{xx}$ and $\sigma_{xy}$ estimated from the Airy stress function and those obtained from the simulations with $\varphi=0.80$, $D=10d$, $\Delta=20d$, and $F_{\rm drag}=1.6\times 10^{-2}k_n d$.
This shows that our estimation qualitatively reproduces the anisotropic behaviors, e.g., the positive and negative peak angles of $\sigma_{xy}$ in front of the intruders.

\section{Discussion}\label{sec:discussion}

Let us discuss our results.
First, we consider the magnitude of $M_{\rm Y}$.
As shown in Fig \ref{fig:mu_fy}, $F_{\rm Y}$ is almost constant ($F_{\rm Y} = 9.2\times10^{2}\mu_{\rm b}mg$).
Now, let us estimate this yield force on the bases of a simple argument.
From Fig.\ \ref{fig:displacement}, it might be a good estimation that the intruders push the surrounding particles inside a square region whose size is $\Delta + D=30d$.
In this case, the number of particles $N_{\rm c}$ present inside this region is approximately given by
\begin{equation}
  N_{\rm c} \simeq \frac{(\Delta + D)^2}{\displaystyle \frac{\pi}{8}(1+1.4^2)d^2}\varphi,
\end{equation}
where we assume that the numbers of particles of diameters $d$ and $1.4d$ are the same.
Under this assumption, the frictional force from the surrounding particles becomes
\begin{equation}
  F_{\rm Y} \simeq \frac{N_{\rm c}}{2}\mu_{\rm b}(1+1.4^2)mg 
  \simeq \frac{4}{\pi}(\Delta +D)^2\varphi\mu_{\rm b}mg.
\end{equation}
For our choice of parameters, this approximately becomes $F_{\rm Y}\simeq 9\times 10^2\mu_{\rm b}mg$, which gives a good estimation of the simulation result in Fig.\ \ref{fig:mu_fy}.

Next, let us discuss the stress fields.
Figure \ref{fig:coeff} shows that two coefficients $E$ and $B_1$ are dominant over other coefficients.
Then, let us assume that
\begin{equation}
  E=-\frac{F_{\rm drag}}{\pi d},\ B_1=\frac{F_{\rm drag}}{\pi d},
\end{equation}
as suggested in the previous section.
In this case, the Airy function becomes
\begin{equation}
  \chi = \frac{F_{\rm drag}}{\pi}J(1-\vartheta) \cos\vartheta.
  \label{eq:chi_simple}
\end{equation}
Let us consider this meaning.
The angle $\vartheta$ is related to the angle $\theta$, which is defined as the angle between the vector $(x, y-\Delta/2)$ and the $x$-axis.
The relationship is approximately given by $\theta \sim 1 - \vartheta \sim \pi/2-\vartheta$.
We also introduce the distance $r$ from the center of the upper (bottom) intruder to the point $(x,y)$. 
With the aid of Eq.\ \eqref{eq:xy_alphabeta}, this is approximately given by
\begin{equation}
  r=\sqrt{x^2 + \left( y \mp \frac{\Delta}{2} \right)^2}
  =\pm \frac{\sqrt{\Delta^2 - D^2}}{2 \sinh \varrho}.
  \label{eq:r}
\end{equation}
Substituting Eq.\ \eqref{eq:r} into Eq.\ \eqref{eq:J}, we get
\begin{equation}
  J=\frac{r \sinh\varrho}{\cosh\varrho - \cos\vartheta}.
  \label{eq:J_r}
\end{equation}
From Eq.\ \eqref{eq:J_r}, Eq.\ \eqref{eq:chi} can be approximately estimated as
\begin{equation}
  \chi\sim\frac{F_{\rm drag}}{\pi} r \theta \sin\theta \frac{\sinh\varrho}{\cosh\varrho - \cos\vartheta}.
\end{equation}
In the vicinity of the upper intruder, $\varrho \to \infty$ is satisfied.
This means that $\sinh\varrho/(\cosh\varrho-\cos\vartheta)\to 1$, and then, we get
\begin{equation}
  \chi\to \frac{F_{\rm drag}}{\pi} r \theta \sin\theta,
\end{equation}
which is consistent with the case when the concentration force is applied to a semi-infinite plane \cite{Timoshenko}.
\begin{figure}
	\centering
	\includegraphics[width=0.85\linewidth]{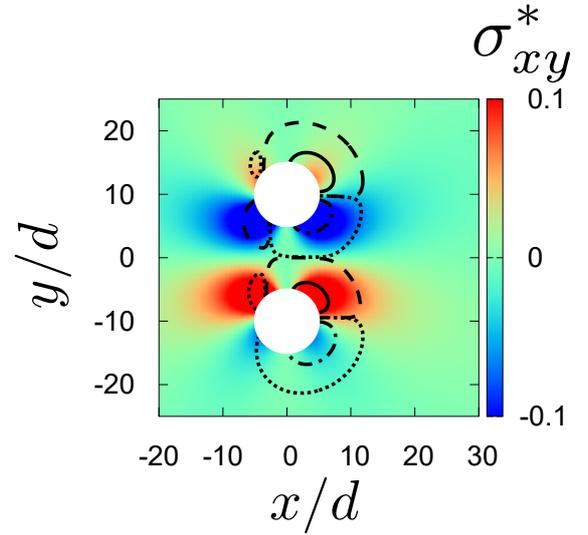}
	\caption{(Color online) Spatial distribution of $\sigma_{xy}$ when we adopt the Airy stress function in Eq.\ \eqref{eq:chi_simple}.
 	The set of parameters is the same as that in Fig.\ \ref{fig:stress_fields}.
 	The solid, dashed, dotted, and dot--dashed lines correspond to the contours of $3.0\times10^{-2}$, $1.0\times 10^{-4}$, $-1.0\times 10^{-4}$, and $-3.0\times10^{-2}$, respectively.}
	\label{fig:stress_simple}
\end{figure}
Figure \ref{fig:stress_simple} shows the stress field $\sigma_{xy}$ obtained from the Airy stress function \eqref{eq:chi_simple} when we choose the same set of parameters as that in Fig.\ \ref{fig:stress_fields}.
This figure indicates that the above simple treatment works as the zeroth order approximation, while the magnitude is underestimated compared with that obtained from the simulation.

\begin{figure}
	\centering
	\includegraphics[width=\linewidth]{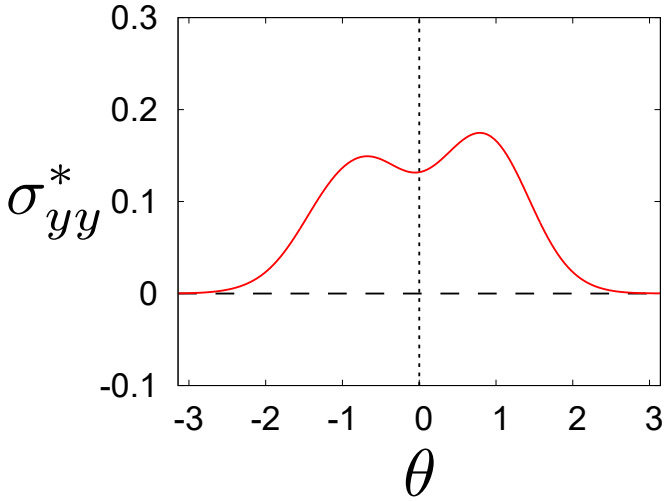}
	\caption{(Color online) Stress field $\sigma_{yy}$ on the surface of the upper intruder for $\varphi=0.80$, $D=10d$, $\Delta=20d$, and $F_{\rm drag}=1.6\times10^{-2}k_nd$.}
	\label{fig:sigma_yy}
\end{figure}

In this study, we do not consider the motions of the intruders in the $y$-direction for simplicity of discussion.
In some previous studies \cite{Dhiman20,Escalante17,Cruz16}, forces perpendicular to the drag direction were investigated, and it was reported that the force between the intruders is basically attractive, but becomes repulsive owing to particle clogging if the distance between the intruders is too narrow ($\Delta-D\gtrsim d$).
In our system, let us check the profile of the stress on the surfaces of the intruders (see Fig.\ \ref{fig:sigma_yy}), where the stress $\sigma_{yy}$ outside the intruders is larger than that inside.
Integrating the stress around the surface, we can understand that the force acting on the intruder in the $y$-direction becomes attractive.
This is consistent with the results of previous studies \cite{Dhiman20,Escalante17,Cruz16}.
More details should be studied in the future.

\section{Summary}\label{sec:summary}
We investigated the drag of two cylindrical objects in a two-dimensional granular environment by the discrete element method.
A yield force was found to exist, and this yield force was insensitive to the distance between the intruders.
When the drag force was above the yield force, we found that the steady speed of the intruders increased as the distance between intruders decreased, because particles between them were influenced by both of them.
On the other hand, the stress fields around the intruders were investigated below the yield force, where the Airy stress function in the bipolar coordinates well reproduced the stress fields obtained from the simulations.


\begin{acknowledgment}
One of the authors (ST) thanks Hisao Hayakawa for fruitful discussions.
This work is financially supported by Grants-in-Aid of MEXT for Scientific Research (Grant Nos.\ JP20K14428 and JP21H01006).
\end{acknowledgment}

\appendix
\section{Determination of System Size}\label{sec:def_L}
In this Appendix, we explain how the system size ($L_x$ and $L_y$) is determined in the simulations.
Because we put the smaller particles at $y=\pm L_y/2$, the system size in the $x$-direction must be an integer multiple of the diameter $d$.
We also note that the aspect ratio of the system should be close to a certain value $\nu$.
Under these conditions, let us determine $L_x$ and $L_y$ for a given packing fraction $\varphi$.

The areas occupied by the intruders, the bulk particles, and the wall particles are, respectively, given by
\begin{subequations}
\begin{align}
  S_{\rm intruder} &= \frac{\pi}{4}D^2 \cdot N_{\rm intruder},\label{eq:S_tracer}\\
  S_{\rm bulk} &= \frac{1}{2}\frac{\pi}{4}(N-N_{\rm wall})(1+1.4^2)d^2,\\
  S_{\rm wall} &= \frac{1}{2}\frac{\pi}{4}d^2N_{\rm wall},\label{eq:S_wall}
\end{align}
\end{subequations}
where $N_{\rm intruder}$ and $N_{\rm wall}$ are the numbers of intruders and wall particles, respectively.
Here, the factor $1/2$ in Eq.\ \eqref{eq:S_wall} comes from the fact that half of the particles are located inside the system.
Summing up Eqs.\ \eqref{eq:S_tracer}--\eqref{eq:S_wall}, the total area occupied by the intruders, the bulk particles, and the wall particles is given by
\begin{equation}
  S
  =\frac{\pi}{4} N_{\rm intruder} D^2+\frac{\pi}{8}(1+1.4^2)N d^2 
  -\frac{\pi}{8}1.4^2 N_{\rm wall}d^2.
  \label{eq:S_total}
\end{equation}

If the system size is given by $L\times \nu L$, the size $L$ is determined by Eq.\ \eqref{eq:S_total} as
\begin{equation}
  \varphi \nu L^2
  =\frac{\pi}{4} N_{\rm intruder}D^2+\frac{\pi}{8}(1+1.4^2)N d^2 
  -\frac{\pi}{8}1.4^2 N_{\rm wall}d^2.
  \label{eq:L_equation}
\end{equation}
Because the wall is made of the smaller particles, $N_{\rm wall}$ should be given by
\begin{equation}
  N_{\rm wall} = 2\left\lfloor \frac{L}{d} \right\rfloor,
  \label{eq:N_wall}
\end{equation}
where $\lfloor x \rfloor$ is the floor function that returns an integer $n$ for $x-1<n\le x$.
In this case, the system area is unchanged when we choose
\begin{equation}
  L_x = \lfloor L \rfloor,\quad
  L_y = \frac{\nu L^2}{\lfloor L \rfloor}.
  \label{eq:LxLy}
\end{equation}
Rigorously, the aspect ratio is larger than $\nu$, but it is very close to $\nu$.
Substituting Eq.\ \eqref{eq:N_wall} into Eq.\ \eqref{eq:L_equation}, we get
\begin{equation}
  L^2 
  +\frac{\pi}{4\varphi\nu}1.4^2d^2\left\lfloor \frac{L}{d} \right\rfloor 
  -\frac{\pi}{4\varphi\nu}N_{\rm intruder}D^2
  -\frac{\pi}{8\varphi\nu}(1+1.4^2)d^2 N = 0.
\end{equation}
We can numerically determine $L$ from this equation.
For our choice of parameters ($N_{\rm intruder}=2$, $N=4000$, $\varphi=0.80$, and $\nu=3/2$), we obtain $L/d\simeq 62.65\cdots$, and then $L_x=62d$ and $L_y=94.96d$.

\section{Case of three intruders}\label{sec:three_intruders}
In this Appendix, we investigate how the drag changes when there exist three intruders in the system.

\begin{figure}
	\centering
	\includegraphics[width=\linewidth]{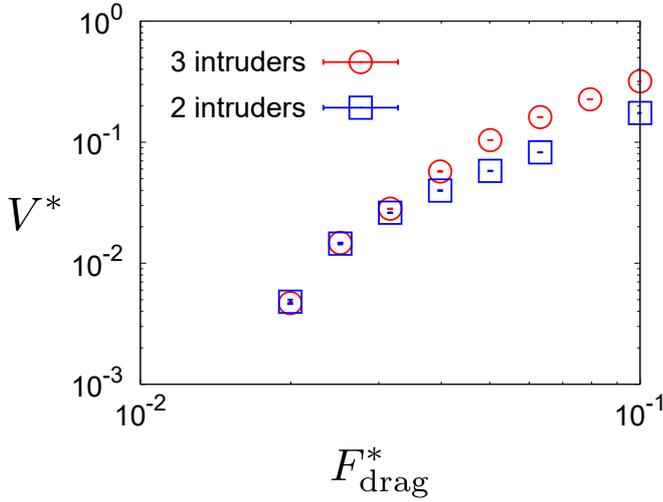}
	\caption{(Color online) Steady speed of the intruders against the drag force in the cases of two (open squares) and three intruders (open circles) for $\varphi = 0.80$, $D = 10d$, and $\Delta = 20d$.}
	\label{fig:FV_2_3}
\end{figure}

First, we discuss the drag law when the drag force is larger than the yield force.
Figure \ref{fig:FV_2_3} shows the drag for two and three intruders.
This suggests that the yield force is almost unchanged because the areas of interactions with the intruders are limited to near them.
On the other hand, the difference becomes larger as the steady speed of the intruders increases.
This can be understood as follows.
As the speed increases, the number of particles that interact with the intruders also increases.
Then, there exist areas where particles interact with two adjacent intruders.
This area becomes more movable than other areas.
Therefore, the resistance from the surrounding particles decreases.
Here, a more systematic study is needed but is beyond the scope of this paper.

\section{Derivation of Stress Fields in terms of Bipolar Coordinates}\label{sec:bipolar}
In this Appendix, we revisit the general solutions of the stress fields around intruders when we adopt bipolar coordinates \cite{Jeffery21}.
The bipolar coordinates $(\varrho, \vartheta)$ are transformed from the $(x, y)$--plane via the following conformal mapping:
\begin{equation}
  x+iy = ia \coth \left(\frac{\varrho+i\vartheta}{2}\right),
  \label{eq:conformal_map}
\end{equation}
with an imaginary number $i$ and a constant $a=\sqrt{\Delta^2-D^2}/2$.
Here, the explicit relationship between $(\varrho, \vartheta)$ and $(x, y)$ is given by
\begin{equation}
\begin{cases}
	\displaystyle x = \frac{a\sin\vartheta}{\cosh\varrho - \cos\vartheta}\\
	\displaystyle y = \frac{a\sinh\varrho}{\cosh\varrho - \cos\vartheta}
\end{cases}.
	\label{eq:xy_alphabeta}
\end{equation}
We note that $\varrho={\rm const.}$ and $\vartheta={\rm const.}$ correspond to the following circles in the $(x, y)$--plane, respectively:
\begin{subequations}\label{eq:xy_alphabeta_2}
\begin{align}
	x^2 + \left(y-a\coth \varrho\right)^2 &= \frac{a^2}{\sinh^2\varrho},\\
	\left(x-a\cot\vartheta\right)^2 + y^2 &= \frac{a^2}{\sin^2\vartheta}.
\end{align}
\end{subequations}
For later convenience, we introduce a quantity $J$ as
\begin{equation}
	J = \sqrt{\left(\frac{\partial x}{\partial \varrho}\right)^2 + \left(\frac{\partial x}{\partial \vartheta}\right)^2}
	= \frac{a}{\cosh\varrho - \cos\vartheta},
\end{equation}
which corresponds to the growth rate of the conformal mapping \eqref{eq:conformal_map}.

Now, let us consider the Airy stress function $\chi$.
In the $(x,y)$--plane, the stress function should satisfy
\begin{equation}
	\Delta \Delta \chi(x,y)=0\label{eq:chi_xy}
\end{equation}
with the Laplacian $\Delta\equiv (\partial^2/\partial x^2+\partial^2/\partial y^2)$.
By using the relationship \eqref{eq:xy_alphabeta}, Eq.\ \eqref{eq:chi_xy} in the $(\varrho,\vartheta)$--plane becomes
\begin{align}
	\left(\frac{\partial^4}{\partial \varrho^4}+2\frac{\partial^4}{\partial \varrho^2\partial \vartheta^2}
	+\frac{\partial^4}{\partial \vartheta^4} -2\frac{\partial^2}{\partial \varrho^2} 
	+ 2\frac{\partial^2}{\partial \vartheta^2}+1\right) \left(\frac{\chi}{J}\right)=0.
	\label{eq:eq_chi_J}
\end{align}
The general solution of this equation is given by
\begin{align}
	\frac{\chi}{J}
	&= (E\cos\vartheta + F \sin\vartheta + G\cosh\varrho + H\sinh\varrho)\vartheta\nonumber\\
		&\hspace{1em}
		+ (B_0\cosh\varrho + D_0\sinh\varrho)\varrho\nonumber\\
		&\hspace{1em}
		+ [A_1 \cosh(2\varrho) + B_1 + C_1 \sinh(2\varrho) + D_1\varrho]\cos\vartheta\nonumber\\
		&\hspace{1em}
		+ [A_1^\prime \cosh(2\varrho) + C_1^\prime \sinh(2\varrho) + D_1^\prime\varrho]\sin\vartheta\nonumber\\
		&\hspace{1em}
		+ \sum_{n=2}^\infty
			\left\{A_n \cosh[(n+1)\varrho] + B_n \cosh[(n-1)\varrho]\right.\nonumber\\
		&\hspace{2em}\left.
				+C_n \sinh[(n+1)\varrho] + D_n\sinh[(n-1)\varrho]\right\}\cos(n\vartheta)\nonumber\\
		&\hspace{1em}
		+ \sum_{n=2}^\infty
			\left\{A_n^\prime \cosh[(n+1)\varrho] + B_n^\prime \cosh[(n-1)\varrho]\right.\nonumber\\
		&\hspace{2em}\left.
				+C_n^\prime \sinh[(n+1)\varrho] + D_n^\prime\sinh[(n-1)\varrho]\right\}\sin(n\vartheta),
	\label{eq:chi/J}
\end{align}
with constants $E$, $F$, $G$, $H$, $A_n$, $B_n$, $C_n$, $D_n$, $A_n^\prime$, $B_n^\prime$, $C_n^\prime$, and $D_n^\prime$ ($n\ge0$).
Using Eq.\ \eqref{eq:chi/J}, the stress can be calculated as \cite{Jeffery21}
\begin{subequations}
\begin{align}
	a\sigma_{\varrho\varrho}
	&= \left[(\cosh\varrho-\cos\vartheta)\frac{\partial^2}{\partial \vartheta^2} 
		-\sinh\varrho \frac{\partial}{\partial \varrho} \right.\nonumber\\
    &\hspace{2em}\left.-\sin\vartheta\frac{\partial}{\partial \vartheta}
		+\cosh\varrho\right]\left(\frac{\chi}{J}\right)\\
	a\sigma_{\vartheta\vartheta}
	&= \left[(\cosh\varrho-\cos\vartheta)\frac{\partial^2}{\partial \varrho^2} 
		-\sinh\varrho \frac{\partial}{\partial \varrho} \right.\nonumber\\
    &\hspace{2em}\left.-\sin\vartheta\frac{\partial}{\partial \vartheta}
		+\cos \vartheta\right]\left(\frac{\chi}{J}\right)\\
	a\sigma_{\varrho\vartheta}
	&= -(\cosh\varrho-\cos\vartheta)\frac{\partial^2}{\partial \varrho \partial \vartheta}\left(\frac{\chi}{J}\right).
\end{align}
\end{subequations}
These are the general solutions of the stress fields.
Once we get the stresses $\sigma_{\varrho\varrho}$, $\sigma_{\varrho\vartheta}$, and $\sigma_{\vartheta\vartheta}$, the stresses on Cartesian coordinates $\sigma_{xx}$, $\sigma_{xy}$, and $\sigma_{yy}$ can be calculated by
\begin{subequations}
\begin{align}
  \sigma_{xx} 
  &= \sigma_{\varrho\varrho}\cos^2\gamma + \sigma_{\vartheta\vartheta}\sin^2\gamma 
  - 2\sigma_{\varrho\vartheta}\sin\gamma\cos\gamma,\\
  \sigma_{xy} 
  &= (\sigma_{\varrho\varrho} - \sigma_{\vartheta\vartheta})\sin\gamma\cos\gamma 
  + \sigma_{\varrho\vartheta}(\cos^2\gamma-\sin^2\gamma),\\
  \sigma_{yy} 
  &= \sigma_{\varrho\varrho}\sin^2\gamma + \sigma_{\vartheta\vartheta}\cos^2\gamma 
  + 2\sigma_{\varrho\vartheta}\sin\gamma\cos\gamma,
\end{align}
\end{subequations}
(see, e.g., Ref.\ \citenum{Timoshenko})
where $\gamma$ satisfies \cite{Jeffery21}
\begin{subequations}
\begin{align}
  \cos\gamma &= -\frac{\sinh\varrho\sin\vartheta}{\cosh\varrho-\cos\vartheta},\\
  \sin\gamma &= -\frac{\cosh\varrho\cos\vartheta-1}{\cosh\varrho-\cos\vartheta}.
\end{align}
\end{subequations}



\begin{thebibliography}{99}

\bibitem{Lamb}
	S.\ H.\ Lamb,
	\textit{Hydrodynamics}
	(Dover, New York, 1945).
\bibitem{Batchelor}
	G.\ K.\ Batchelor, 
	\textit{An Introduction to Fluid Dynamics}
	(Cambridge Univ.\ Press, Cambridge, 1967).
\bibitem{Takada20jet}
 	S.\ Takada and H.\ Hayakawa,
	Phys.\ Rev.\ Res.\ \textbf{2}, 033468 (2020).
\bibitem{Hilton13}
	J.\ E.\ Hilton and A.\ Tordesillas,
	Phys.\ Rev.\ E \textbf{88}, 062203  (2013).
\bibitem{Takada17_3D}
	S.\ Kumar, K.\ A.\ Reddy, S.\ Takada, and H.\ Hayakawa, 
	arXiv:1712.09057.
\bibitem{Takada20}
	S.\ Takada and H.\ Hayakawa,
	Granul.\ Matter \textbf{22}, 6 (2020).
\bibitem{Dhiman20}
	M.\ Dhiman, S.\ Kumar, K.\ A.\ Reddy, and R.\ Gupta,
	J.\ Fluid Mech.\ \textbf{886},\ A23 (2020).
\bibitem{Takehara10}
	Y.\ Takehara, S.\ Fujimoto, and K.\ Okumura, 
	EPL \textbf{92}, 44003 (2010).
\bibitem{Takehara14}
	Y.\ Takehara and K.\ Okumura,
	Phys.\ Rev.\ Lett.\ \textbf{112}, 148001 (2014).
\bibitem{Takada17}
	S.\ Takada and H.\ Hayakawa,
	J.\ Eng.\ Mech.\ \textbf{143}, C4016004 (2017).
\bibitem{Wassgren03}
	C.\ R.\ Wassgren and J.\ A.\ Cordova,
	Phys.\ Fluids \textbf{15}, 3318 (2003).
\bibitem{Reddy11}
	K.\ A.\ Reddy, Y.\ Forterre, and O.\ Pouliquen,
	Phys.\ Rev.\ Lett.\ \textbf{106}, 108301 (2011).
\bibitem{Guillard14}
    F.\ Guillard, Y.\ Forterre, and O.\ Pouliquen, 
	Phys.\ Fluids \textbf{26}, 043301 (2014).
\bibitem{Pal21}
    A.\ Pal and A.\ Kudrolli,
    Phys.\ Rev.\ Fluids \textbf{6}, 124302 (2021).
\bibitem{Kubota21}
	T.\ Kubota, H.\ Ishikawa, and S.\ Takada,
	EPJ Web Conf.\ \textbf{249}, 03033 (2021).
\bibitem{Cheng07}
	X.\ Cheng, G.\ Varas, D.\ Citron, H.\ M.\ Jaeger, and S.\ R.\ Nagel,
	Phys.\ Rev.\ Lett.\ \textbf{99}, 188001 (2007).
\bibitem{Cheng14}
 	X.\ Cheng, L.\ Gordillo, W.\ W.\ Zhang, H.\ M.\ Jaeger, and S.\ R.\ Nagel,
 	Phys.\ Rev.\ E \textbf{89}, 042201 (2014).
\bibitem{Sano12}
	T.\ G.\ Sano and H.\ Hayakawa,
	Phys.\ Rev.\ E \textbf{86}, 041308 (2012).
\bibitem{Sano13}
	T.\ G.\ Sano and H.\ Hayakawa, 
	Prog.\ Theor.\ Exp.\ Phys.\ \textbf{2013}, 103J02 (2013).
\bibitem{Jewel18}
	R.\ Jewel, A.\ Panaitescu, and A.\ Kudrolli,
	Phys. Rev. Fluids \textbf{3}, 084303 (2018).
\bibitem{Cruz16}
	R.\ A.\ L\'{o}pez de la Cruz and G.\ A.\ Caballero-Robledo,
	J.\ Fluid Mech.\ \textbf{800}, 248 (2016).
\bibitem{Escalante17}
	M.\ F.\ Acevedo-Escalante and G.\ A.\ Caballero-Robledo,
	EPJ Web Conf.\ \textbf{140}, 03048 (2017).
\bibitem{Hossain20Mobility}
	T.\ Hossain and P. Rognon,
	Phys.\ Rev.\ E \textbf{102}, 022914 (2020).
\bibitem{Hossain20Rate-dependent}
	T.\ Hossain and P.\ Rognon,
	Granul,\ Matter \textbf{22}, 72 (2020).
\bibitem{Hossain20Drag}
	T.\ Hossain and P.\ Rognon,
	Phys.\ Rev.\ Fluids \textbf{5}, 054306 (2020).
\bibitem{Bharadwaj06}
	R.\ Bharadwaj and C.\ Wassgren,
	Phys.\ Fluids \textbf{18}, 043301 (2006).
\bibitem{Katsuragi}
	H.\ Katsuragi,
	\textit{Physics of Soft Impact and Cratering}
	(Springer, Berlin, 2016).
\bibitem{Espinosa21}
	M.\ Espinosa, V.\ L.\ Díaz, A.\ Serrano, and E.\ Altshuler,
	arXiv:2110.15311.
\bibitem{Cundall}
	P.\ A.\ Cundall and O.\ D.\ L.\ Strack,
	G\'{e}otechnique \textbf{29}, 47 (1979).
\bibitem{Luding08}
	S.\ Luding,
	Granul.\ Matter \textbf{10}, 235 (2008).
\bibitem{Kuninaka01}
	H.\ Kuninaka and H.\ Hayakawa, 
	J.\ Phys.\ Soc.\ Jpn.\ \textbf{70}, 2220 (2001).
\bibitem{Hayakawa02}
	H.\ Hayakawa and H.\ Kuninaka,
	Chem.\ Eng.\ Sci.\ \textbf{57}, 239 (2002).		
\bibitem{Garzo_book}
	V.\ Garz\'{o},
	\textit{Granular Gaseous Flows ---A Kinetic Theory Approach to Granular Gaseous Flows---}
	(Springer, Berlin, 2019).
\bibitem{Zhang10}
	J.\ Zhang, R.\ P.\ Behringer, and I.\ Goldhirsch,
	Prog.\ Theor.\ Phys.\ Suppl.\ \textbf{184}, 16 (2010).
\bibitem{Timoshenko}
	S.\ P.\ Timoshenko and J.\ N.\ Goodier,
	\textit{Theory of Elasticity}
	(McGraw-Hill, New York, 1970) 3rd ed.
\bibitem{Daniels17}
    K.\ E.\ Daniels, J.\ E.\ Kollmer, and J.\ G.\ Puckett,
    Rev.\ Sci.\ Instrum.\ \textbf{88}, 051808 (2017).
\bibitem{Jeffery21}
	G.\ B.\ Jeffery,
	Phil.\ Trans.\ Roy.\ Soc.\ A \textbf{221}, 265 (1921).
\end{thebibliography}
\end{document}